\documentclass[prc,twocolumn,showpacs,amsmath,amssymb]{revtex4}
\usepackage{times}
\usepackage{graphicx}
\usepackage{dcolumn}
\usepackage{bm}
\usepackage{amstext}

\begin{document}

\title{Effects of isovector scalar $\delta$-meson on hypernuclei}
 
\author{M. Ikram$^{1}$, S. K. Singh$^{2}$, S. K. Biswal$^{2}$ and S. K. Patra$^{2}$}
\affiliation{$^1$ Department of Physics, Aligarh Muslim University, Aligarh-202002, India\\ 
$^2$ Institute of Physics, Sachivalaya Marg, Bhubaneswar-751005, India.}


\date{\today}

\begin{abstract}
We analyze the effects of $\delta-$ meson on hypernuclei within the 
frame-work of relativistic mean field theory.
The $\delta-$ meson is included into the Lagrangian for hypernuclei.
The extra nucleon-meson coupling ($g_\delta$) affects the every piece of 
physical observables, like binding energy, radii and single particle 
energy of hypernuclei. 
The lambda mean field potential is investigated which is 
consistent with other predictions.
Flipping of single particle energy levels are observed with the strength of 
$g_\delta$ in the considered hypernuclei as well as normal nuclei.
The spin-orbit potentials are observed for considered hypernuclei and the effect 
of $g_\delta$ on spin-orbit potentials is also analyzed.
The calculated single-$\Lambda$ binding energies ($B_\Lambda$) are 
quite agreeable with the experimental data.

\end{abstract}
\maketitle

\section{Introduction}
Normal nuclei are quite informative for showing the 
distinctive features of nucleon-nucleon (NN) interaction.
The knowledge on NN interaction may be extended to 
hyperon-nucleon (YN) or hyperon-hyperon (YY) interaction 
by injecting one or more strange baryon to bound nuclear 
system~\cite{cugnon2000,gibson1995,rufa1990,rufa1987,
hiyama2010,ehiyama2010,schulze2010,povh1987,chrien1989,
gal2009,schaffner1994,gal2010,gal2004,Hiyama2010,harada2005,lanskoy1998}.
The injected hyperon originates a new quanta of strangeness 
and makes a more interesting nuclear system with increasing 
density~\cite{gibson1995}.
Unlike to nucleons, a hyperon is not Pauli blocked owing to 
strangeness quantum number and resides at the centre of the nucleus.
Hyperons are used as an impurity in nuclear systems to reveal many of the 
nuclear properties in the dimension of strangeness
~\cite{baldo1990,tan2004,lu2011}.
For this, a slightly unbound normal nucleus can be 
bound by addition of $\Lambda$ particle~\cite{vretenar1998,zhou2008}.

To understand the structure of strange system, it is necessary 
to evaluate the contribution of YN interaction. 
But, due to short life-time of hyperon, only limited 
information on YN scattering data is available which is 
a major consequence of the experimental difficulties~\cite{hashimoto2006}.
For this purpose, more theoretical data are needed to 
explore the strangeness physics.
However, extensive efforts on theoretical basis have been made 
to enrich the knowledge about YN interaction using relativistic 
and non-relativistic mean field approaches. 
For example, 
Skyrme Hartree-Fock (SHF)~\cite{rayet1976,rayet1981,cugnon2000,
rshyam2012,schulze2010}, deformed Hartree-Fock 
(DHF)~\cite{xian2007,zhou2008}, Skyrme Hartree-Fock with 
BCS approach~\cite{win2011} and relativistic mean field 
(RMF) formalism~\cite{mares1989,mares1993,glendenning1993,mares1994,
vretenar1998,rufa1990,rufa1987,ren2012,
brockmann1977,boguta1981}.

From last three decades, the relativistic mean field theory reproduces 
the experimental data on binding energy, root mean square (rms) radius, 
and quadrupole deformation parameter for finite nuclei throughout the 
periodic chart~\cite{walecka1974,bodmer1977,walecka1986,serot1992,
pannert1987,ring1996,patra1991,patra2007}.
Here the degrees of freedom are nucleons and mesons. 
To deal with hypernuclei, one has to incorporate the meson-hyperon interaction 
to the relativistic Lagrangian.
The most successful RMF model of Boguta and Bodmer, included the $\sigma-$, 
$\omega-$, and $\rho-$mesons along with the nonlinear coupling of $\sigma-$meson, 
which simulates the three-body interaction~\cite{schiff1951}.
The $\rho-$meson takes care the neutron-proton asymmetry, while the Coulomb 
interaction is taken care by the electromagnetic field produced by the protons.
Although, conventional RMF model is quite successful, but it is recently 
realized that the isovector-scalar $\delta-$meson, which arises from 
the mass and isospin asymmetry of proton and neutron is very 
important for nuclear system with much difference 
in neutron N and proton Z number~\cite{kubis1997,roca2011,singh2013}. 
The main objective of the present study is to see the effects of 
$\delta-$meson for some selected hypernuclear systems.
For this purpose, we evaluate the contribution of $\delta-$meson in 
hypernuclear system and make a comparison with normal nuclei.

The paper is organized as follows: Section II gives a brief 
description of relativistic mean field formalism for hypernuclei 
with inclusion of $\delta-$ meson.
The results are presented and discussed in Section III.
Selection of $g_\delta$ and $g_\rho$ coupling constant is 
also discussed in this section.
The paper is summarized in Section IV.

\section{Formalism }
The RMF Lagrangian for hyperon-nucleon-meson many-body system including the 
$\delta-$ meson is written as~\cite{rufa1990,rufa1987,glendenning1993,mares1994,
sugahara1994,vretenar1998,win2008,lu2003}:

\begin{eqnarray} 
\mathcal{L}&=&\mathcal{L}_N+\mathcal{L}_\Lambda \;, 
\end{eqnarray}

\begin{eqnarray}
{\cal L}_N&=&\bar{\psi_{i}}\{i\gamma^{\mu}
\partial_{\mu}-M\}\psi_{i}
+{\frac12}(\partial^{\mu}\sigma\partial_{\mu}\sigma
-m_{\sigma}^{2}\sigma^{2})		   \nonumber \\
&-&{\frac13}g_{2}\sigma^{3} 
-{\frac14}g_{3}\sigma^{4}
+{\frac12}(\partial^{\mu}\delta\partial_{\mu}\delta
-m_{\delta}^{2}\delta^{2})                      \nonumber \\
&-&g_{s}\bar{\psi_{i}}\psi_{i}\sigma 		
-g_{\delta}\bar{\psi_{i}}\vec{\tau}\psi_{i}\vec\delta 
-{\frac14}\Omega^{\mu\nu}\Omega_{\mu\nu}
+{\frac12}m_{w}^{2}V^{\mu}V_{\mu}		\nonumber \\
&-&g_{w}\bar\psi_{i}\gamma^{\mu}\psi_{i}V_{\mu}    
-{\frac14}B^{\mu\nu}B_{\mu\nu} 
+{\frac12}m_{\rho}^{2}{\vec{R}^{\mu}}{\vec{R}_{\mu}}  \nonumber \\
&-&{\frac14}F^{\mu\nu}F_{\mu\nu}                    
-g_{\rho}\bar\psi_{i}\gamma^{\mu}\vec{\tau}\psi_{i}\vec{R^{\mu}} \nonumber \\
&-&e\bar\psi_{i}\gamma^{\mu}\frac{\left(1-\tau_{3i}\right)}{2}\psi_{i}A_{\mu} \;,   \\
\mathcal{L}_{\Lambda}&=&\bar\psi_\Lambda\{i\gamma^\mu\partial_\mu
-m_\Lambda\}\psi_\Lambda
-g_{\omega\Lambda}\bar\psi_\Lambda\gamma^{\mu}\psi_\Lambda V_\mu   \nonumber \\
&-&g_{\sigma\Lambda}\bar\psi_\Lambda\psi_\Lambda\sigma \;,
\end{eqnarray}
where $\psi$ and $\psi_\Lambda$ denote the Dirac spinors for 
nucleon and $\Lambda$ particle, whose masses are M and
$m_\Lambda$ respectively, and $g_{\sigma\Lambda}$, 
$g_{\omega\Lambda}$ are $\Lambda-$meson coupling constants.
Because of zero isospin, the $\Lambda$ hyperon does not couple 
to $\rho-$ and $\delta-$mesons.
The quantities $m_{\sigma}$, $m_{\omega}$, $m_{\rho}$ and $m_{\delta}$ are the 
masses for $\sigma-$, $\omega-$, $\rho-$ and $\delta-$mesons. 
The field for the ${\sigma}$-meson is denoted by ${\sigma}-$, 
${\omega}-$meson by $V_{\mu}$, ${\rho}-$meson by $R_{\mu}$ 
and $\delta$-meson by $\delta$. 
The quantities $g_s$, $g_{\omega}$, $g_{\rho}$, $g_{\delta}$ and $e^2/4{\pi}$=1/137 
are the coupling constants for the ${\sigma}-$, ${\omega}-$, ${\rho}-$, 
${\delta}-$mesons and photon, respectively.
We have $g_2$ and $g_3$ self-interaction coupling constants 
for ${\sigma}-$mesons.
The field tensors of the vector, isovector mesons and of the 
electromagnetic field are given by
\begin{eqnarray} 
\Omega^{\mu\nu}& =& \partial^{\mu} V^{\nu} - \partial^{\nu} V^{\mu} \;,\nonumber \\
B^{\mu\nu}& =& \partial^{\mu}R^{\nu} - \partial^{\nu}R^{\mu}\;,  \nonumber \\
F^{\mu\nu}& =& \partial^{\mu}A^{\nu} - \partial^{\nu}A^{\mu} \;.
\end{eqnarray}
The classical variational principle is used to solve the 
field equations for bosons and Fermions.
The Dirac equation for the nucleon is written as:
\begin{equation} 
[-i\alpha.\nabla + V(r)+\beta(M+S(r))]\psi_i=\epsilon_i\psi_i\; ,
\end{equation}
where V(r) and S(r) represent the vector and scalar potential, defined as 
\begin{equation} 
V(r)=g_{\omega}V_{0}(r)+g_{\rho}\tau_{3}\rho_{0}(r)+e\frac{(1-\tau_3)}{2}A_0(r)\; ,
\end{equation}
and
\begin{equation} 
S(r)=g_{\sigma}\sigma(r)+\tau_{3}g_{\delta}\delta_0(r) \;,
\end{equation}
where subscript $i=$ n, p for neutron and proton, respectively.
The Dirac equation for $\Lambda$ particle is 
\begin{equation} 
[-i\alpha.\nabla + \beta\big(m_\Lambda+g_{\sigma\Lambda}\sigma(r)\big)+
g_{\omega\Lambda}V_0(r)]\psi_\Lambda = \epsilon_{\Lambda}\psi_\Lambda \;.
\end{equation}
The field equations for bosons are
\begin{eqnarray} 
\{-\bigtriangleup+m^2_\sigma\}\sigma(r) &=& -g_\sigma\rho_s(r)-g_2\sigma^2(r)
-g_3\sigma^3(r) \nonumber \\
&&- g_{\sigma\Lambda}\rho_s^{\Lambda}(r)\; ,                             \nonumber \\ 
\{-\bigtriangleup+m^2_\omega\}V_0(r) &=& g_\omega\rho_v(r)+
g_{\omega\Lambda} \rho_v^{\Lambda}(r)\; ,                                 \nonumber \\
\{-\bigtriangleup+m^2_\delta\}\delta_3(r) &=& -g_\delta\rho^s_3(r) \;,   \nonumber \\
\{-\bigtriangleup+m^2_\rho\}R^0_3(r) &=& g_\rho\rho_3(r) \;,             \nonumber \\
-\bigtriangleup A_0(r) &=& e\rho_c(r)\; .
\end{eqnarray}
Here $\rho_s$, $\rho_s^{\Lambda}$, $\rho_v$ and $\rho_v^{\Lambda}$ are 
the scalar and vector density for $\sigma-$ and $\omega-$field 
in nuclear and hypernuclear system which are expressed as
\begin{eqnarray} 
\rho_s(r) &=& \sum_ {i=n,p}\bar\psi_i(r)\psi_i(r)\;,                         \nonumber \\
\rho_s^{\Lambda}(r) &=& \sum \bar\psi_{\Lambda}(r)\psi_\Lambda(r)\;,           \nonumber \\
\rho_v(r) &=& \sum_{i=n,p}\psi^\dag_i(r)\psi_i(r) \;,			        \nonumber \\
\rho_v^{\Lambda}(r) &=& \sum \psi^\dag_{\Lambda}(r)\psi_\Lambda(r)\;.            
\end{eqnarray}
The scalar density for $\delta-$ field is 
\begin{eqnarray} 
\rho^s_3(r) &=& \sum_{i=n,p} \bar\psi_i(r)\tau_{3i}\psi_i(r)\;.
\end{eqnarray}
The vector density $\rho_3(r)$ for $\rho-$field and charge density 
$\rho_c(r)$ are expressed by 
\begin{eqnarray} 
\rho_3(r) &=& \sum_{i=n,p} \psi^\dag_i(r)\gamma^0\tau_{3i}\psi_i(r)\; ,		      \nonumber \\
\rho_c(r) &=& \sum_{i=n,p} \psi^\dag_i(r)\gamma^0\frac{(1-\tau_{3i})}{2}\psi_i(r)\;.    
\end{eqnarray}
The various rms radii are defined as
\begin{eqnarray}
\langle r_p^2\rangle &=& \frac{1}{Z}\int r_p^{2}d^{3}r\rho_p\;,        \nonumber \\
\langle r_n^2\rangle &=& \frac{1}{N}\int r_n^{2}d^{3}r\rho_n\;,        \nonumber \\
\langle r_m^2\rangle &=& \frac{1}{A}\int r_m^{2}d^{3}r\rho\;,           \nonumber \\
\langle r_\Lambda^2\rangle &=& \frac{1}{\Lambda}\int r_\Lambda^{2}d^{3}r\rho_\Lambda\;, 
\end{eqnarray}
for proton, neutron, matter and lambda rms radii respectively and 
$\rho_p$, $\rho_n$, $\rho$ and $\rho_\Lambda$ are their corresponding densities. 
The charge rms radius can be found from the proton rms radius 
using the relation $r_{c} = \sqrt{r_p^2+0.64}$ taking into 
consideration the finite size of the proton.
The total energy of the system is given by 
\begin{eqnarray}
E_{total} &=& E_{part}(N,\Lambda)+E_{\sigma}+E_{\omega}+E_{\delta}+E_{\rho}	\nonumber \\
&&+E_{c}+E_{pair}+E_{c.m.},   
\end{eqnarray}
where $E_{part}(N,\Lambda)$ is the sum of the single particle energies of the 
nucleons (N) and hyperon ($\Lambda$).
$E_{\sigma}$, $E_{\omega}$, $E_{\delta}$, $E_{\rho}$, $E_{c}$, 
$E_{pair}$ and $E_{cm}$ are the contributions of meson fields, 
Coulomb field, pairing energy and the center-of-mass energy, respectively.
We use NL3* parameter set through out the calculations~\cite{lalazissis09}. 

We adopt the relative $\sigma-$ and $\omega-$ coupling to find the 
numerical values of $\Lambda-$meson coupling constants.
The relative coupling constants for $\sigma$ and $\omega$ field are defined as 
$R_\sigma=g_{\sigma\Lambda}/g_\sigma$ and $R_\omega=g_{\omega\Lambda}/g_\omega$.
We use the value of the relative $\omega$ coupling as $R_\omega=2/3$ from 
the naive quark model~\cite{keil2000,shen2006}.   
For used NL3* parameter set, we take the relative $\sigma$ coupling value 
as $R_\sigma=0.620$~\cite{ren2012}.
In present calculations, to take care of pairing interaction the 
constant gap BCS approximation is used and the centre of mass 
correction is included by the formula $E_{cm}=-(3/4)41A^{-1/3}$.


\section{RESULTS AND DISCUSSIONS}

The calculated results are shown in Table~\ref{tab1} and Figs. (1$-$11) 
for both normal nuclei and hypernuclei. 
We study the effect of $\delta-$meson on some selected hypernuclei, like 
$^{48}_\Lambda$Ca, $^{90}_\Lambda$Zr and $^{208}_\Lambda$Pb.
To demonstrate the effect of $g_\delta$ on hypernuclei, we make a comparison with 
their normal nuclear ($^{48}$Ca, $^{90}$Zr and $^{208}$Pb) counter parts.

\begin{table*}
\caption{\label{tab1}The calculated lambda binding energy, $B_\Lambda$ 
for single-$\Lambda$ hypernuclei is compared with the experimental 
data~\cite{qnusmani1991,hashimoto2006,rshyam2012}, given in brackets. 
The used parameter set is pure NL3* without any inclusion of $g_\delta$. 
The radii are also displayed. Energies are given in MeV and radii are in fm.}
\renewcommand{\arraystretch}{1.5}
\begin{ruledtabular}
\begin{tabular}{lcccccccc}
&BE (MeV) &$B_\Lambda$(s) &$B_\Lambda$(p) &$r_c$ &$r_t$  &$r_p$ &$r_n$ &$r_\Lambda$ \\
\hline
$^{16}_\Lambda$N     &130.04 &13.80 (13.76$\pm$0.16)&3.56  (2.84$\pm$0.16)&2.563  &2.468   &2.442   &2.510 &2.302\\
$^{16}_\Lambda$O     &126.44 &13.80 (12.5$\pm$0.35) &3.56  (2.5$\pm$0.5)  &2.666  &2.476   &2.544   &2.420 &2.301\\
$^{28}_\Lambda$Si    &235.94 &20.18 (16.6$\pm$0.2)  &9.10  (7.0$\pm$1.0)  &2.991  &2.826   &2.883   &2.800 &2.323\\
$^{32}_\Lambda$S     &273.74 &21.77 (17.5$\pm$0.5)  &10.20 (8.1$\pm$0.6)  &3.158  &2.982   &3.056   &2.943 &2.287\\
$^{40}_\Lambda$Ca    &346.91 &20.03 (18.7$\pm$1.1)  &11.24 (11.0$\pm$0.6) &3.436  &3.286   &3.343   &3.258 &2.612\\
$^{51}_\Lambda$V     &456.05 &22.10 (19.97$\pm$0.13)&13.88 (11.28$\pm$0.6)&3.547  &3.490   &3.461   &3.540 &2.735\\
$^{89}_\Lambda$Y     &790.09 &24.19 (23.1$\pm$0.5)  &17.78 (16.0$\pm$1.0) &4.216  &4.204   &4.145   &4.270 &3.130\\
$^{139}_\Lambda$La   &1186.67&25.18 (24.5$\pm$1.2)  &20.49 (20.1$\pm$0.4) &4.835  &4.895   &4.776   &4.991 &3.657\\
$^{208}_\Lambda$Pb   &1659.77&26.58 (26.3$\pm$0.8)  &22.67 (21.3$\pm$0.7) &5.490  &5.602   &5.439   &5.718 &4.017\\

\end{tabular}
\end{ruledtabular}
\end{table*}

\subsection{Strategy to fit $g_\rho$ and $g_\delta$:}

The NL3* parametrization used in RMF is fitted phenomenologically.
All the masses and their coupling constants are adjusted 
to reproduce some specific experimental data. 
Therefore, it is not just to add one more parameter like 
$g_{\delta}$, to study it's effect keeping all other parameters 
of NL3* as fixed.  
It might be possible that the physics described by $g_{\delta}$ 
may already be inbuilt in the sub parameters of NL3* and the inclusion 
of $\delta-$meson coupling may lead towards a double counting. 

In this regard, we might expect a connection between $g_{\delta}$ 
and $g_{\rho}$ since both the coupling constants are isospin dependent. 
In such a situation, there are two possible ways 
for this problem to avoid the double counting: 
(i) to consider a dependency on both $g_{\delta}$ and $g_{\rho}$ couplings.  
In this case, modify the parameter $g_{\rho}$ to 
fit an experimental data which is linked to both
$g_{\rho}$ and $g_{\delta}$ for each new given value of $g_{\delta}$, 
such as binding energy or (ii) to get a completely new parameter set 
including this interaction to consider as a new degree of freedom from the 
beginning, i.e., start from an {\it ab initio} 
calculations as done in Ref.~\cite{hofmann01}. 

Here, we are not interested to make a new parameter by 
inclusion of this interaction but our motive is just to 
extract the contribution of $\delta-$meson in hypernuclei 
and corresponding normal nuclei. 
For this, we adopt the first approach to 
analyze the effect of $g_\delta$ on hypernuclei. 
The combination of $g_\delta$ and $g_\rho$ are chose 
in such a way that for a given value of $g_\delta$, 
the combined contribution of $g_\delta$ and $g_\rho$ 
(by adjusting $g_\rho$) reproduces the physical observable 
which exactly match with the original predictions when 
$g_\delta$ was not included. 
We implement this scheme on binding energy which is the 
best physical observables to see every effects in the nuclear system. 
So, we choose the binding energies of 
$^{48}_\Lambda$Ca, $^{90}_\Lambda$Zr, $^{208}_\Lambda$Pb hypernuclei 
and corresponding their normal nuclei to consider as an experimental data. 
By inclusion of $g_\delta$, the binding energies  
change from their original predictions.
To bring back the NL3* binding energies for considered nuclei 
and hypernuclei, we modifiy the $g_{\rho}$ coupling. 
In this way, we get various combinations of 
($g_\rho$, $g_\delta$) for different given values of $g_\delta$. 
As we have already mentioned, the combinations of $g_\delta$ and $g_\rho$
are possible because of both of the coupling constants are linked with isospin. 

\subsection{Binding energy, radii and single particle energy}
Before going to task on $\delta-$meson, it is necessary to 
check the reliability of the parameter which is going to be used.
For this purpose, we calculate the total binding energy (BE), 
single lambda binding energy ($B_\Lambda$) and radii for some selected 
hypernuclei whose experimental data are available. 
After analyzing Table~\ref{tab1}, we found that the lambda 
binding energy $B_\Lambda$ (for s- and p-state) are quite 
comparable with the experimental data.
For example, the $B_\Lambda$ of $^{16}_\Lambda$N is 13.8 MeV 
in our calculation, and the experimental value is (13.76$\pm$0.16) MeV.
It is obvious that the $\Lambda$ hyperon exhibits its strange 
behaviour and enhance the binding of nucleons in hypernucleus.
The other thing is, with increasing the mass number the lambda 
density becomes smaller in respect to nucleon density and 
as a result lambda radius ($r_\Lambda$) grows up. 
This observation is reflected in Table~\ref{tab1}, where 
$r_\Lambda$ increases with increasing the nuclear number.

In this section, we analyze the effects of $\delta-$meson 
on considered hypernuclei and make their comparison with normal 
nuclei to demonstrate the affects, which is the central theme of the paper. 
For the same, we calculate the binding energy (BE), root mean 
square neutron ($r_n$), proton ($r_p$), charge ($r_{ch}$) 
and matter radius ($r_{t}$), and energy of first and last 
filled orbitals of $^{48}_\Lambda$Ca, $^{90}_\Lambda$Zr, 
$^{208}_\Lambda$Pb and $^{48}$Ca, $^{90}$Zr, $^{208}$Pb 
with various combinations of $g_\rho$ and $g_{\delta}$.

\begin{figure}
\vspace{0.6cm}
\includegraphics[width=1.0\columnwidth]{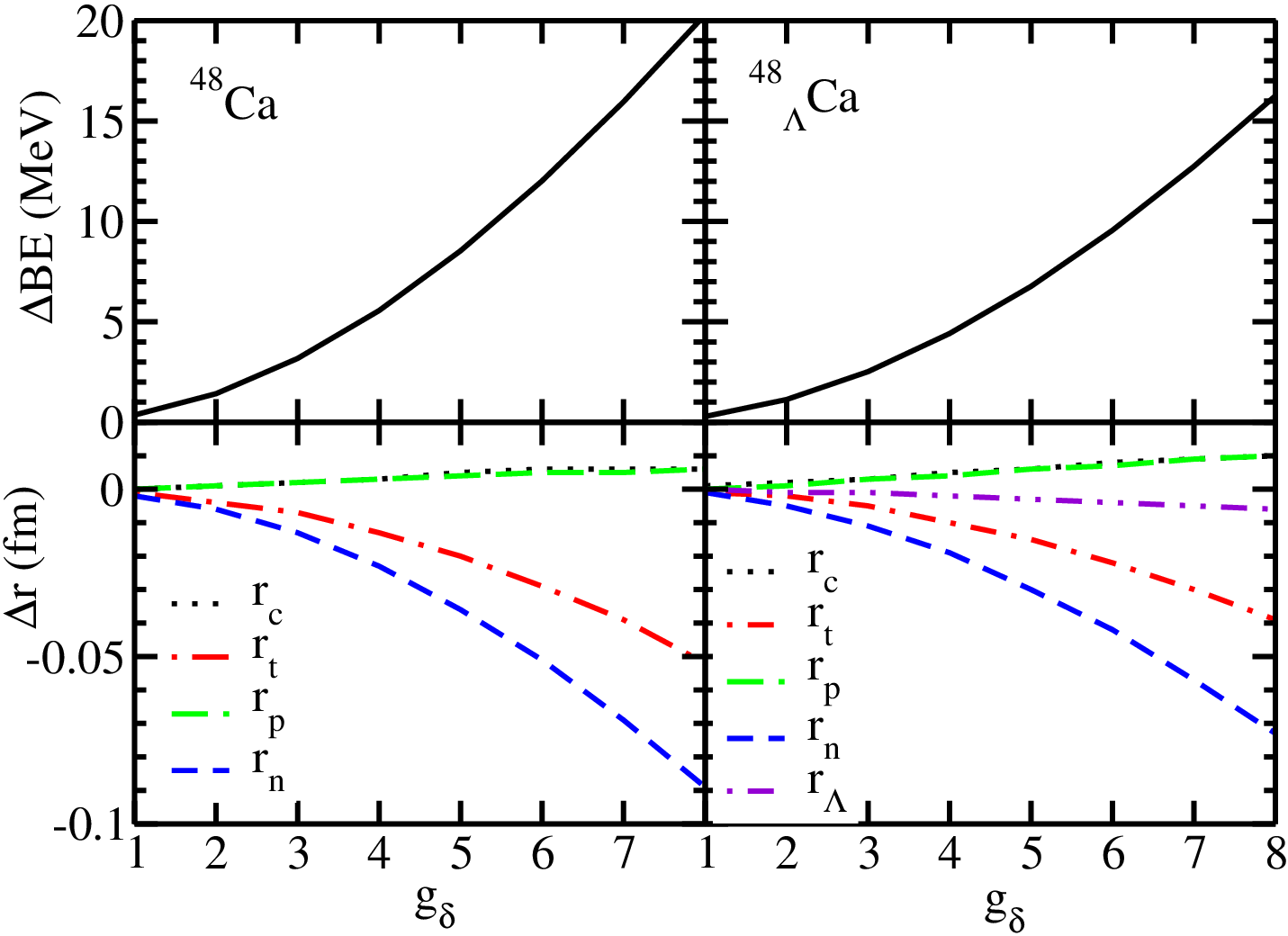}
\caption{\label{fig1}(Color online) Binding energy (BE) and root 
mean square radius for $^{48}$Ca and $^{48}_\Lambda$Ca using 
various combination of $g_\rho$ and $g_{\delta}$.}  
\end{figure}

\begin{figure}
\vspace{0.6cm}
\includegraphics[width=1.0\columnwidth]{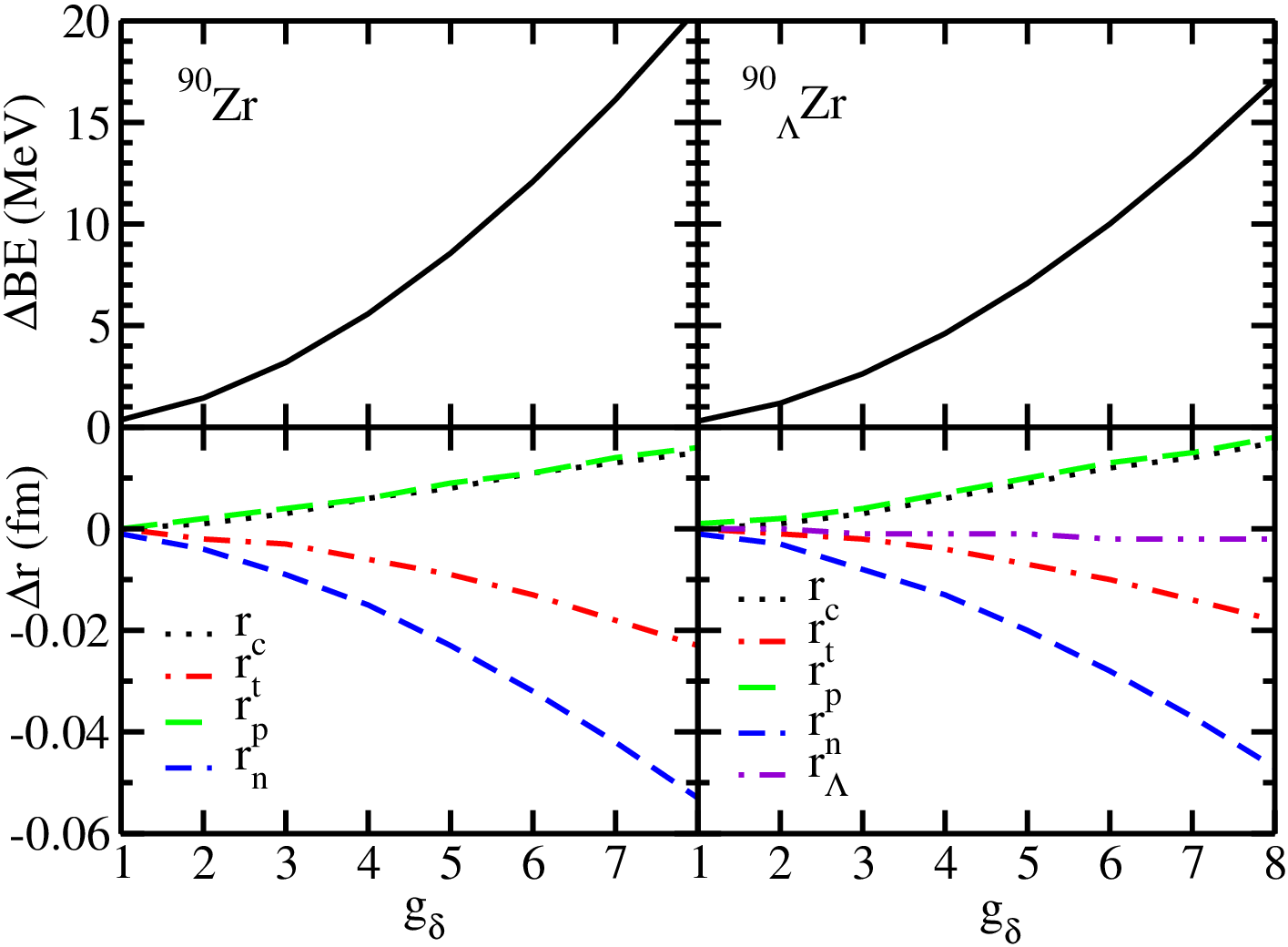}
\caption{\label{fig2}(Color online) same as Fig.1 but 
for $^{90}$Zr and $^{90}_\Lambda$Zr.}  
\end{figure}

\begin{figure}
\vspace{0.6cm}
\includegraphics[width=1.0\columnwidth]{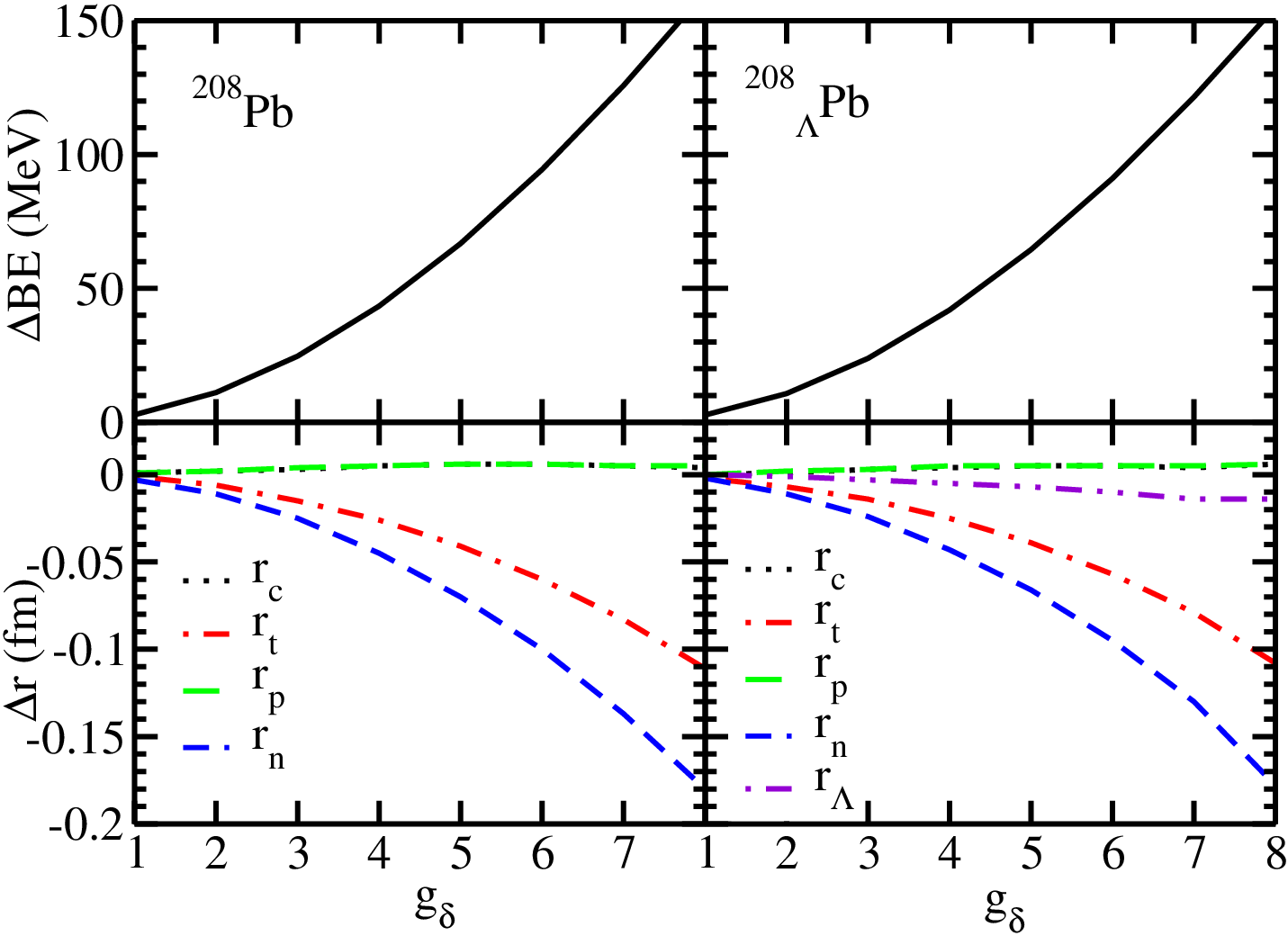}
\caption{\label{fig3}(Color online) same as Fig.1 but 
for $^{208}$Pb and $^{208}_\Lambda$Pb.}  
\end{figure}

\begin{figure}
\vspace{0.6cm}
\includegraphics[width=1.0\columnwidth]{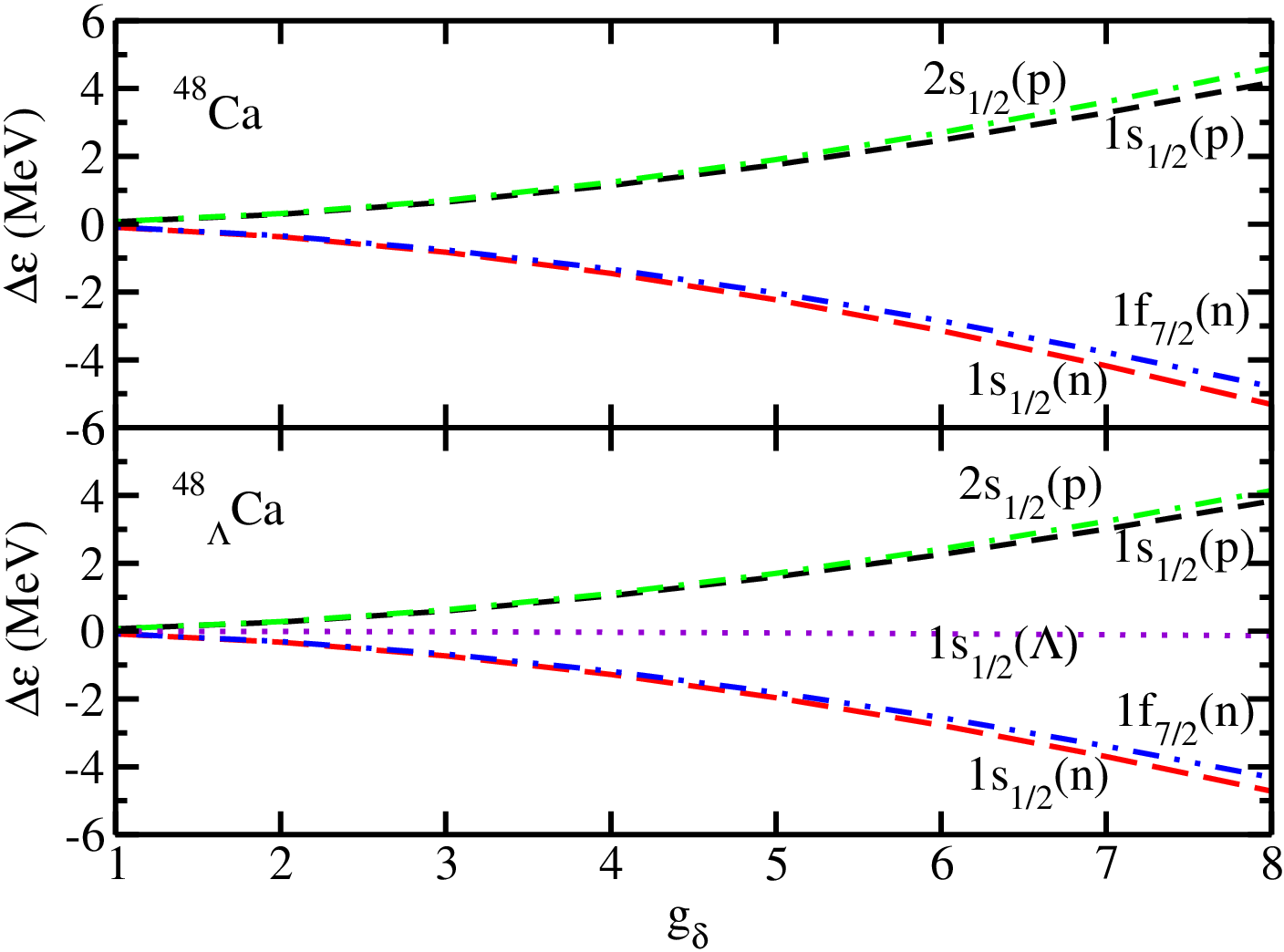}
\caption{\label{fig4}(Color online) First ($1s^{n,p}$) and last 
($1f^n$, $2s^p$) occupied orbits for $^{48}$Ca and $^{48}_\Lambda$Ca 
using various ($g_\rho$, $g_{\delta}$) combinations.}  
\end{figure}

\begin{figure}
\vspace{0.6cm}
\includegraphics[width=1.0\columnwidth]{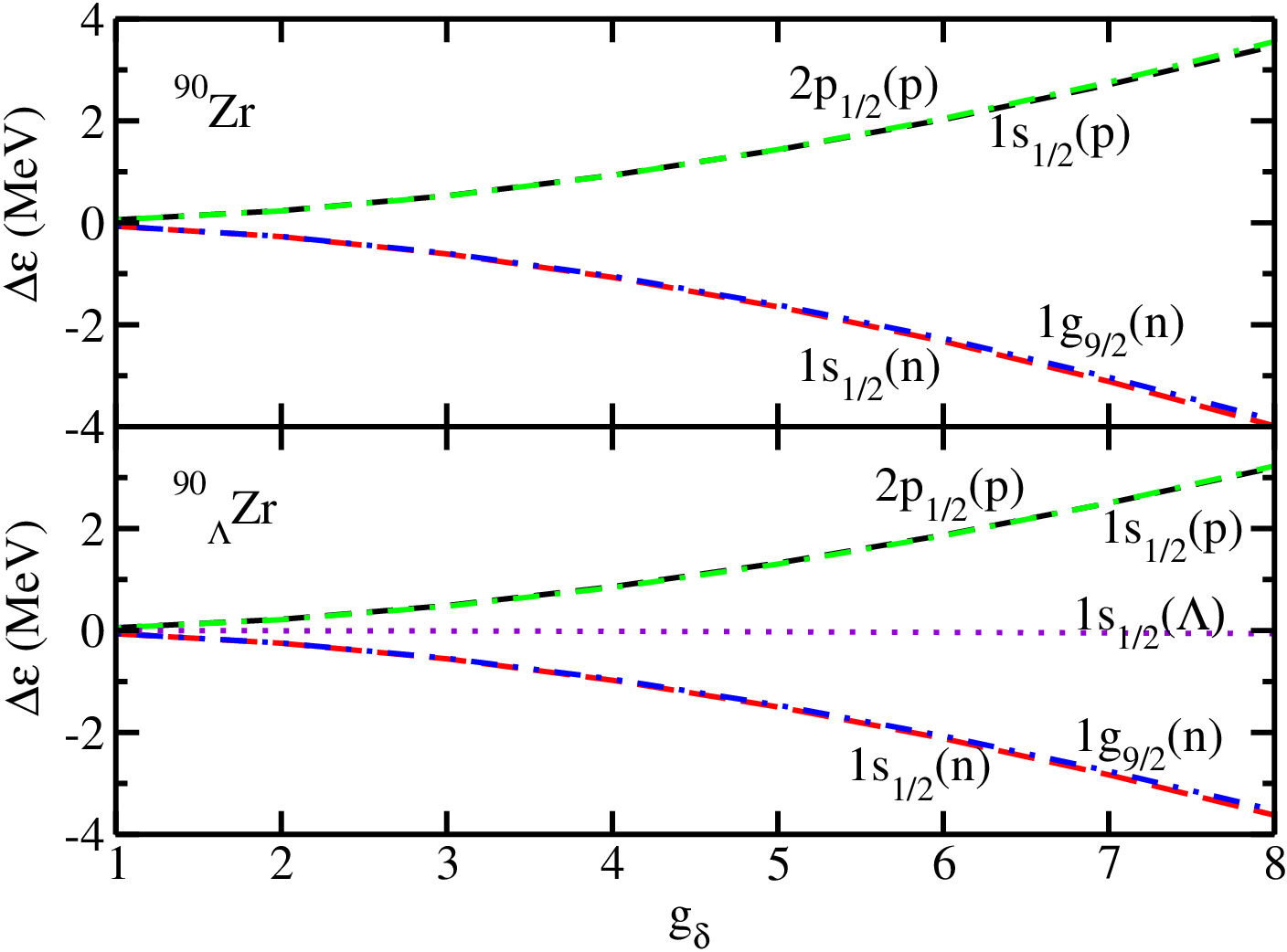}
\caption{\label{fig5}(Color online) First ($1s^{n,p}$) and last 
($1g^n$, $2p^p$) occupied orbits for $^{90}$Zr and $^{90}_\Lambda$Zr 
using various ($g_\rho$, $g_{\delta}$) combinations.}  
\end{figure}

\begin{figure}
\vspace{0.6cm}
\includegraphics[width=1.0\columnwidth]{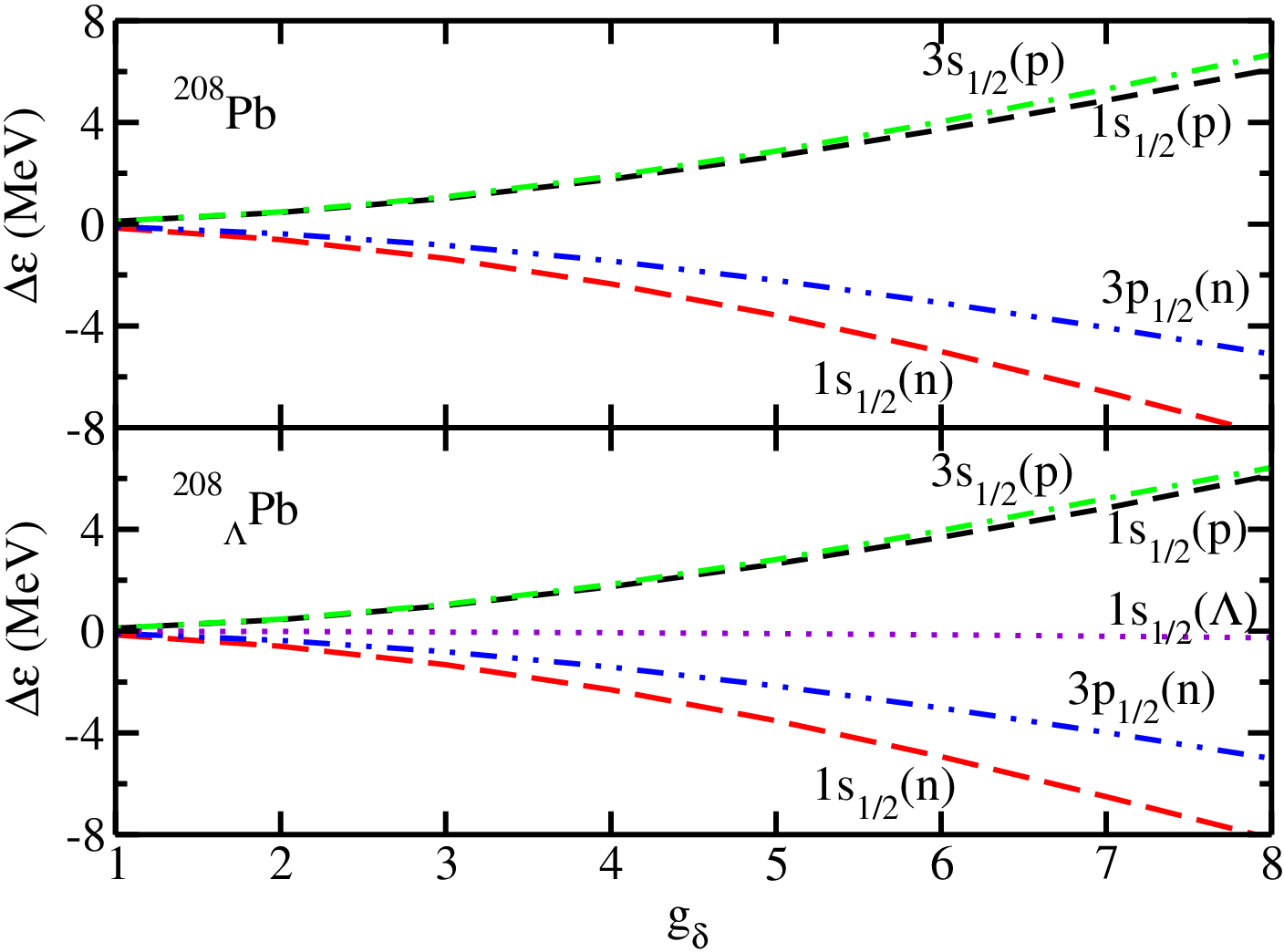}
\caption{\label{fig6}(Color online) First ($1s^{n,p}$) and last 
($3p^n$, $3s^p$) occupied orbits for $^{208}$Pb and $^{208}_\Lambda$Pb 
using various ($g_\rho$, $g_{\delta}$) combinations.}  
\end{figure}

In Fig.~\ref{fig1} (a) and (c), we have shown the binding energy difference 
$\Delta BE$ of $^{48}$Ca and $^{48}_\Lambda$Ca between the two solutions 
obtained with ($g_\rho, g_\delta$=0) and  ($g_\rho, g_\delta$), i.e.  
\begin{equation}
\Delta BE = BE(g_\rho, g_\delta=0) - BE(g_\rho, g_\delta),
\end{equation}
here 
$BE(g_\rho, g_\delta=0)$ is the binding energy at $g_\delta=0$
in the adjusted combination of ($g_\rho, g_\delta$)
and $BE(g_\rho, g_\delta)$ is the binding energy with non-zero 
value of $g_\delta$ in the adjusted combination which reproduce 
the same binding as pure NL3*.
Here, the value of $g_\rho$ used in adjusted combination with 
$g_\delta$ is different from the actual value given in original 
NL3* parameter set.
In other words, we can say that, this strategy evolve 
a new parameter set with extra coupling constant $g_\delta$, 
which also reproduces exactly same physical observables as NL3* set.
Using this procedure, the contribution of $\delta-$ meson 
in binding energy is obtained from $\Delta BE$.
Similarly, the effect of $\delta-$meson in radius for both 
nuclei and hypernuclei can be seen from:
\begin{equation}
\Delta r = r( g_\rho, g_\delta=0) - r( g_\rho, g_\delta),
\end{equation}
where  
$r(g_\rho, g_\delta=0)$ is the 
radius at $g_\delta=0$ in the adjusted combination 
of ($g_\rho, g_\delta$) and $r(g_\rho, g_\delta)$ is the 
radius in adjusted combination of $g_\rho$ and $g_\delta$ 
with non zero value of $g_\delta$, produces exactly 
same experimental value as pure NL3*.
The magnitude of $\Delta r$ with respect to $g_\delta$ for 
$^{48}$Ca, $^{90}$Zr, $^{208}$Pb and their hypernucleus 
$^{48}_\Lambda$Ca, $^{90}_\Lambda$Zr, $^{208}_\Lambda$Pb 
are shown in Figs.~\ref{fig1}$-$\ref{fig3}. 
The same procedure has adopted to estimate the contribution 
of $\delta-$meson on single particle energy for 
considered hypernuclei and their non-strange counter parts, 
which are shown in Figures.~\ref{fig4}$-$\ref{fig6}. 
The difference in single particle energy ($\Delta {\epsilon}$) 
for a particular level is expressed as 
\begin{equation}
\Delta {\epsilon} = {\epsilon}( g_\rho, g_\delta=0) - 
{\epsilon}( g_\rho, g_\delta),
\label{diff_level_d}
\end{equation}
where 
${\epsilon}(g_\rho, g_\delta=0)$ is the single-particle energy
for adjusted combination ($g_\rho$, $g_\delta$) with $g_\delta=0$, 
and ${\epsilon}(g_\rho, g_\delta)$ is energy of the occupied level  
with non zero value of $g_\delta$.

From Figs.\ref{fig1}$-$\ref{fig6}, it is evident 
that the binding energies, radii, single particle energies  
and spin-orbit splitting of nuclei and hypernuclei are 
affected with $g_{\delta}$. 
Because of the presence of $\Lambda$ hyperon, the  contribution of 
$\delta-$meson in binding energies, radii and single particle 
energies are less in hypernuclei compared to normal nuclei.
In other words, we can say that $\delta-$meson affects 
the physical observables less in strange nuclei relative 
to nonstrange nuclei.
In contrary to this, the proton and charge radii are affected 
more in hypernuclei compared to normal nuclei.
From the overview of $g_\delta$ on radii, we find that 
$r_p$ and $r_c$ are in opposite trend with $r_n$, $r_t$, and 
that's why the magnitude of differences of $r_p$ and $r_c$ 
increases with decreasing the asymmetry of the system by addition of hyperon.
A very small reduction on lambda radius is observed with increasing 
strength of $g_\delta$ as shown in Figs.~\ref{fig1}$-$~\ref{fig3}, while the 
lambda potential is completely unaffected by $g_\delta$.
It may happen because of the rearrangement of the levels due to 
presence of lambda particle.
It is to be noticed that there are no convergence solutions 
beyond $g_\delta \sim 8.0$.

In Fig.~\ref{fig4}, we have shown the change in single particle energy
$\Delta{\epsilon_{n,p}}$ of the first ($1s^{n,p}$) and last 
($1f^n$ and $2s^p$) occupied orbitals for $^{48}$Ca, and $^{48}_\Lambda$Ca. 
In the same way, the change in first ($1s^{n,p}$) and last occupied levels 
($1g^n$ and $2p^p$) for $^{90}$Zr and $^{90}_\Lambda$Zr with 
the strength of $g_\delta$ is shown in Fig.~\ref{fig5}.
We also get the same trend in the magnitude of single particle 
energy difference for first ($1s^{n,p}$) and last 
occupied levels ($3p^n$ and $3s^p$) in $^{208}_\Lambda$Pb and 
corresponding their normal nucleus ($^{208}$Pb) which are 
displayed in Fig.~\ref{fig6}.
The magnitude of the difference of single particle energy 
for both neutron and proton (first and last occupied) orbitals 
of considered hypernuclei is small comparable to normal nuclei.
Owing to zero isospin of $\Lambda$ hyperon, the lambda orbit 
($1s_{1/2}^{\Lambda}$) is unaffected with the strength of $g_\delta$. 

\begin{figure}
\vspace{0.6cm}
\includegraphics[width=1.0\columnwidth]{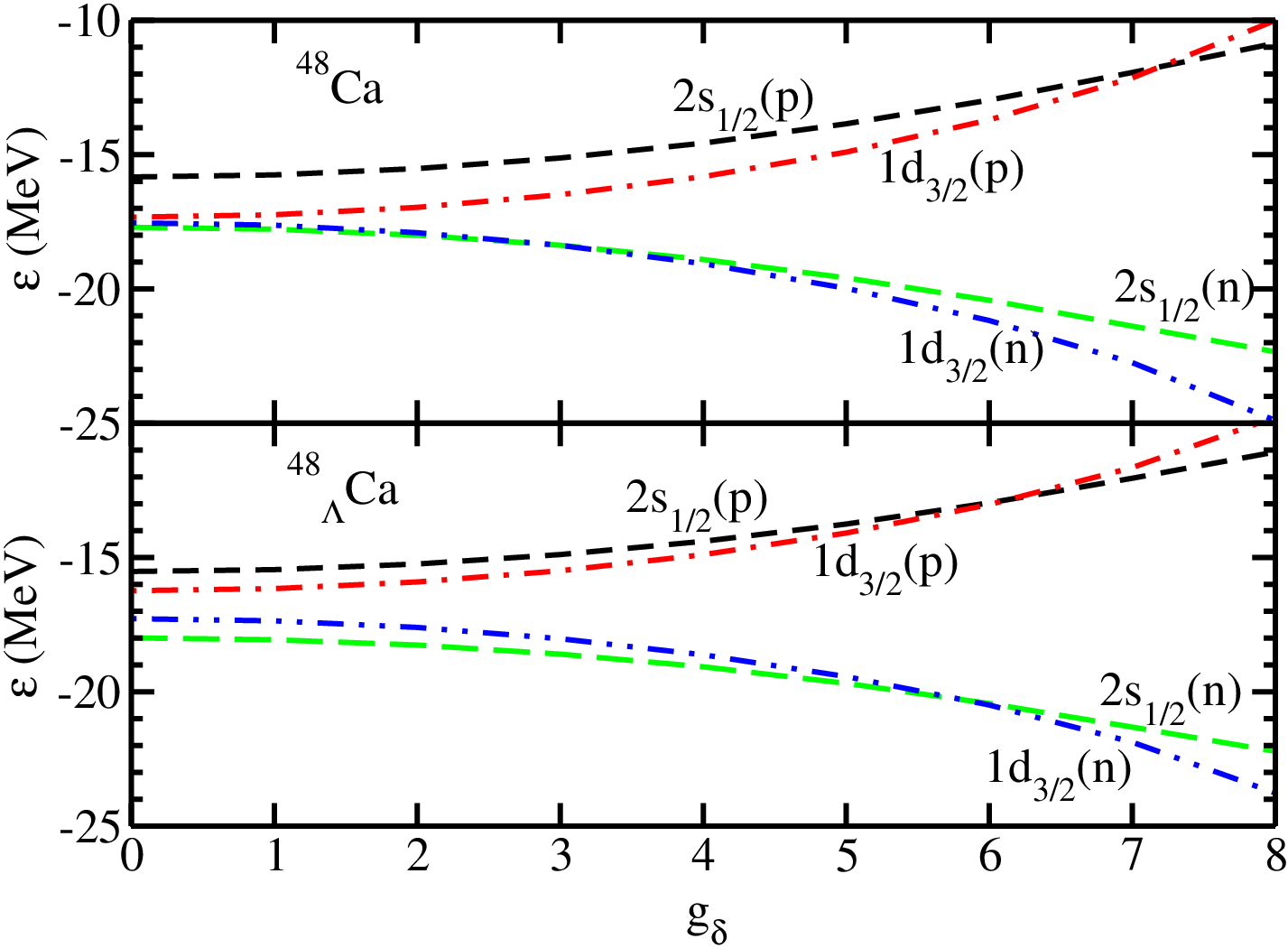}
\caption{\label{fig8}(Color online) The effect of $g_\delta$ on 
spin-orbit splitting is observed for $^{48}$Ca and $^{48}_\Lambda$Ca.}  
\end{figure}

\begin{figure}
\vspace{0.6cm}
\includegraphics[width=1.0\columnwidth]{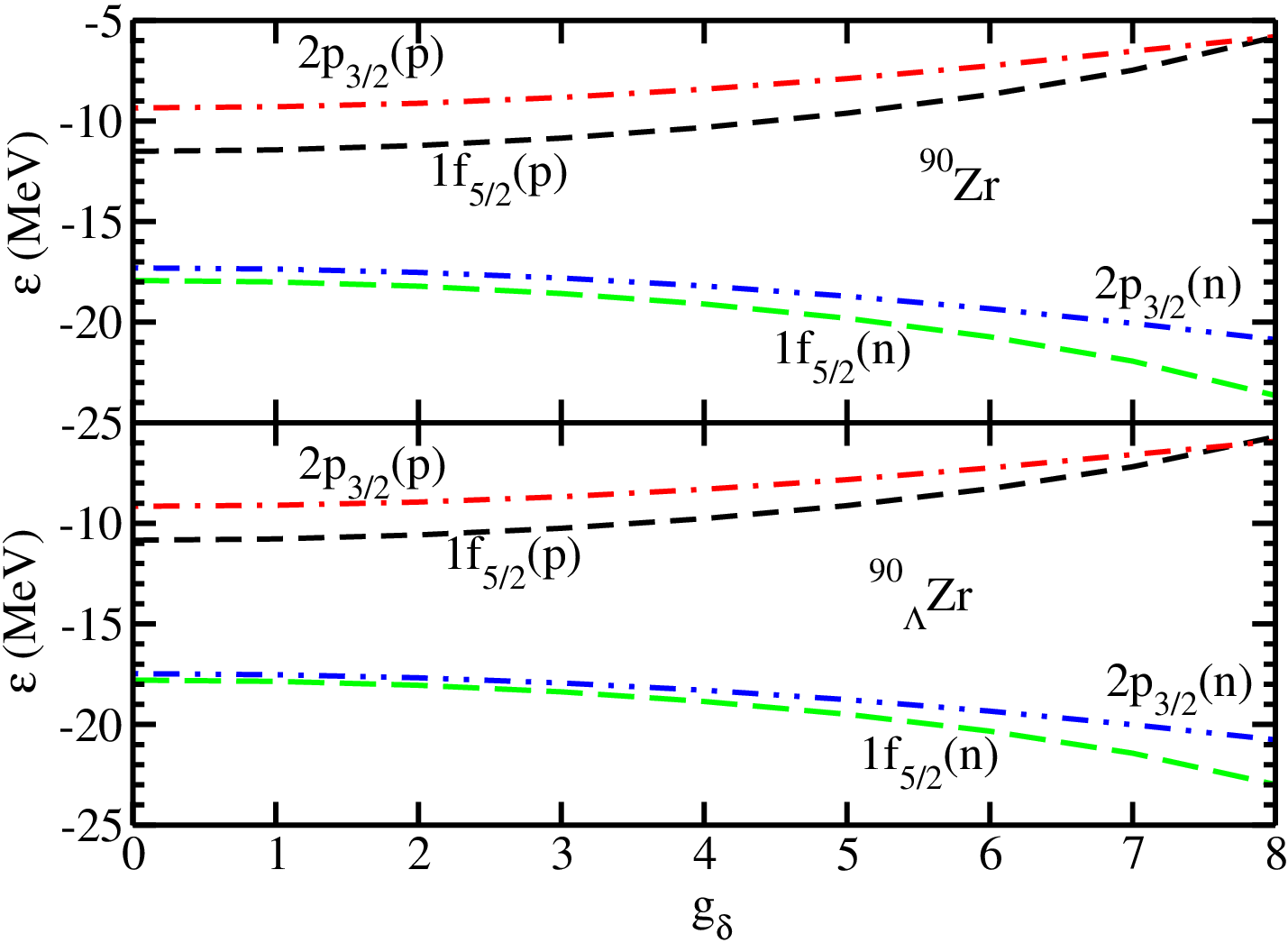}
\caption{\label{fig9}(Color online) same as Fig.7 but for $^{90}$Zr 
and $^{90}_\Lambda$Zr.}  
\end{figure}

\begin{figure}
\vspace{0.6cm}
\includegraphics[width=1.0\columnwidth]{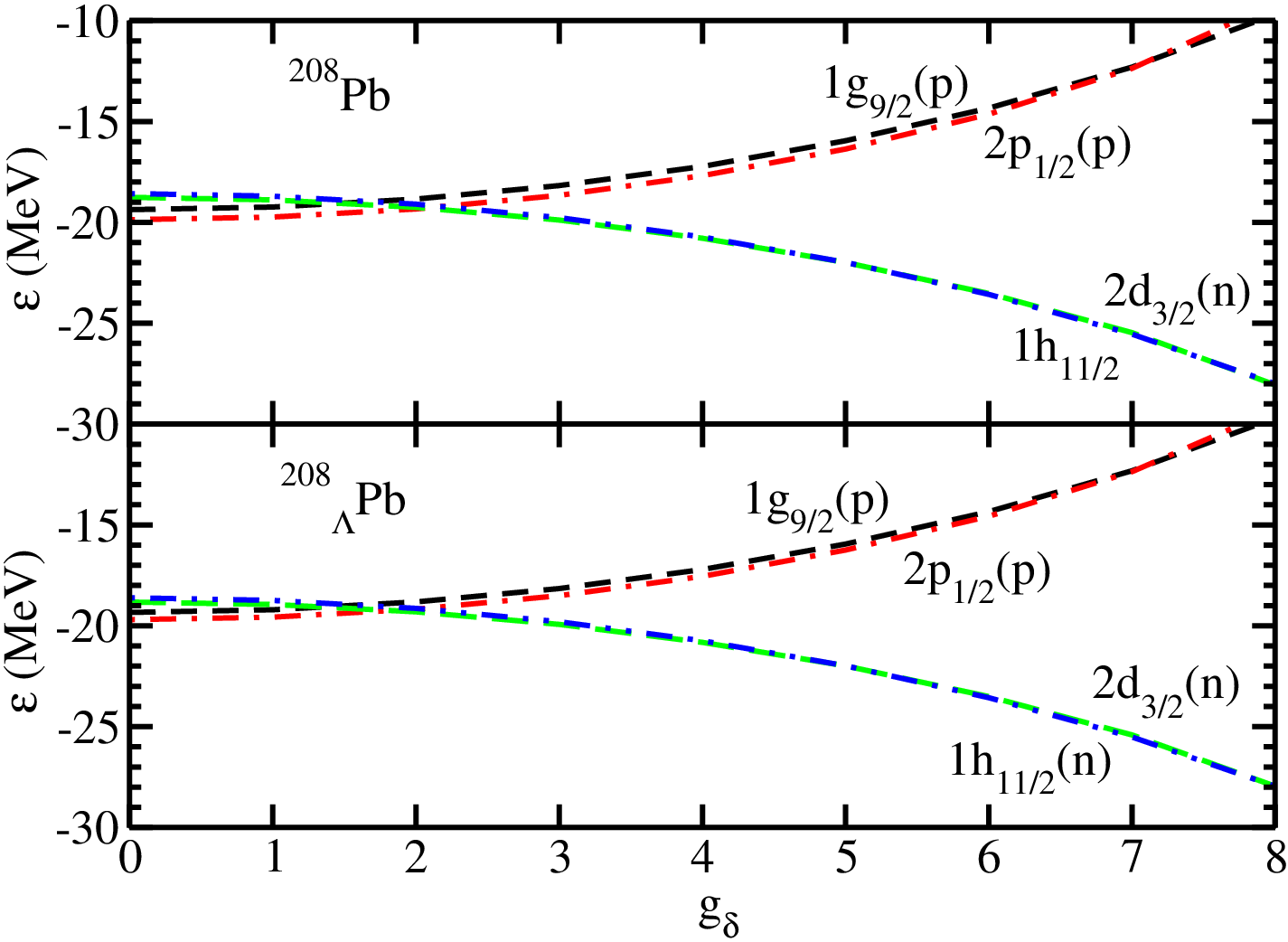}
\caption{\label{fig10}(Color online) same as Fig.7 but for $^{208}$Pb 
and $^{208}_\Lambda$Pb.}  
\end{figure}

\begin{figure}
\vspace{0.6cm}
\includegraphics[width=1.0\columnwidth]{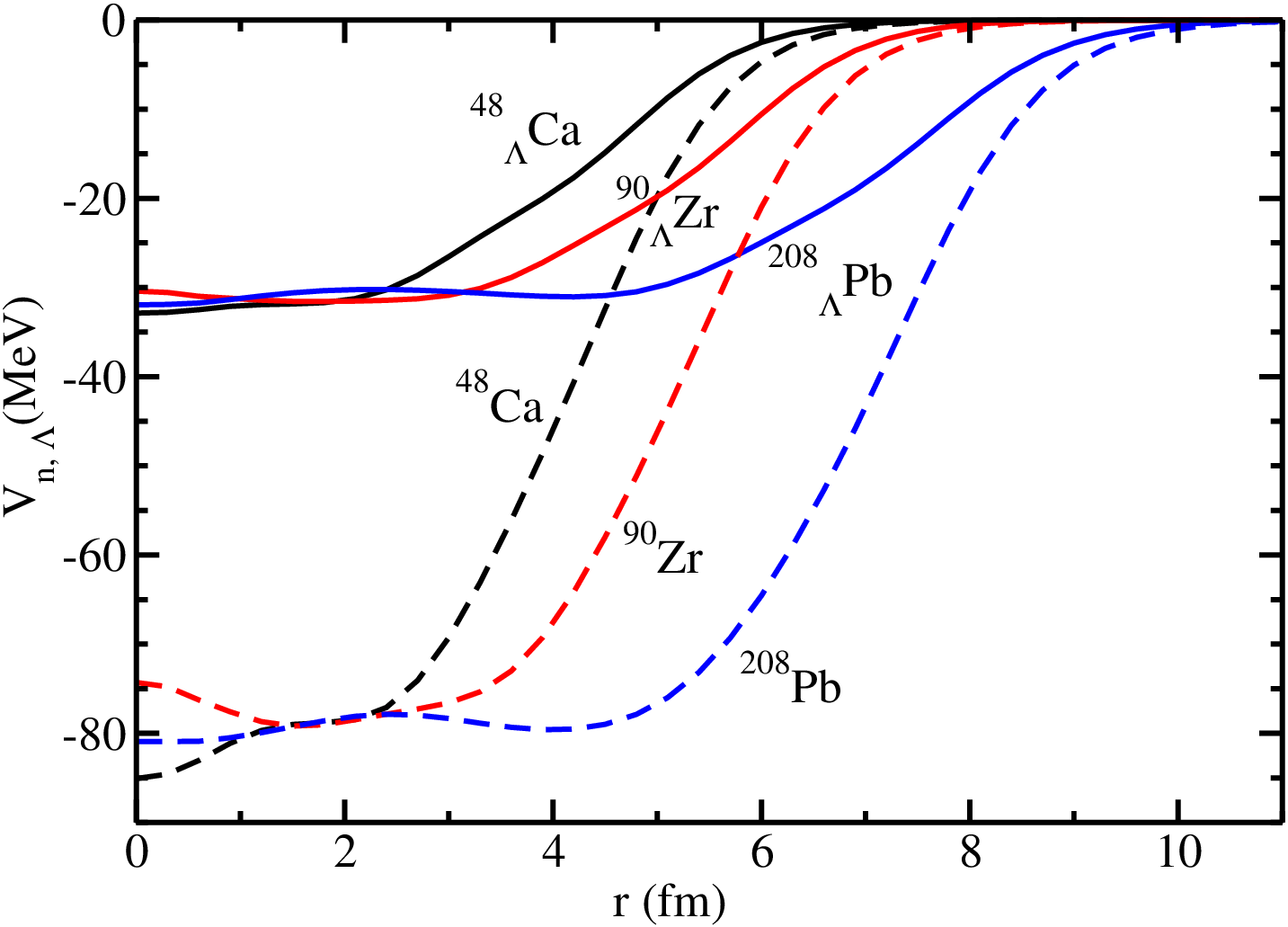}
\caption{\label{fig7}(Color online) The neutron 
($V_n=V_\sigma+V_\omega+V_\rho$) and lambda 
($V_\Lambda=V_{\sigma\Lambda}+V_{\omega\Lambda}$) mean 
field potentials are plotted for $^{48}_\Lambda$Ca, 
$^{90}_\Lambda$Zr and $^{208}_\Lambda$Pb hypernuclei. 
The used parameter set is pure NL3*.}  
\end{figure}

\begin{figure}
\vspace{0.6cm}
\includegraphics[width=1.0\columnwidth]{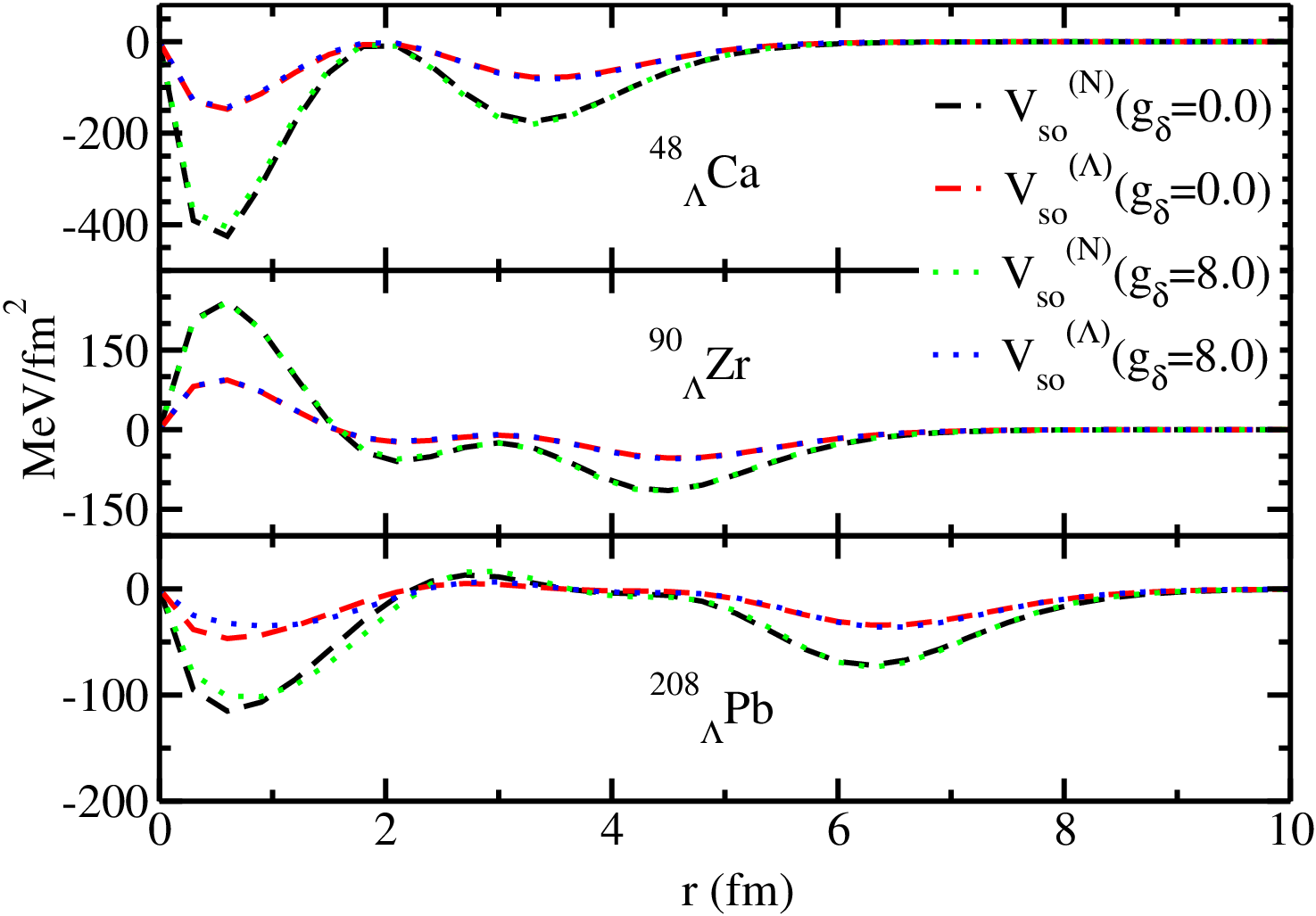}
\caption{\label{fig11}(Color online) Radial dependence of 
spin-orbit potential ($V_{so}^N$ and $V_{so}^{\Lambda}$) 
are plotted for $^{48}_\Lambda$Ca, $^{90}_\Lambda$Zr and 
$^{208}_\Lambda$Pb hypernuclei for different strength of 
$g_\delta$.}  
\end{figure}

After analyzing the single particle spectra for both nuclei and 
hypernuclei, we notice that the orbitals make a shift with 
the strength of $g_\delta$.
In case of $^{48}$Ca, the $2s_{1/2}^{n,p}$ levels are flipped with 
$1d_{3/2}^{n,p}$ in hypernucleus and normal nucleus also.
It is shown in Fig.~\ref{fig9}, the $^{90}$Zr spectra  pretend the 
flipping between $2p_{3/2}^{p}$ and $1f_{5/2}^{p}$ levels for 
both strange and nonstrange nuclei, 
however the strength is low, while the same orbitals 
($2p_{3/2}^{n}$ and $1f_{5/2}^{n}$) for neutron goes 
apart from each other with increasing the strength of $g_\delta$. 
The same trend as $^{48}_\Lambda$Ca is observed for 
$^{208}_\Lambda$Pb and its normal nucleus as shown in Fig~\ref{fig10}.
The proton level $1g_{9/2}^p$ close to flip with $2p_{1/2}^p$, and  
the neutron levels ($2d_{3/2}^n$ and $1h_{11/2}^n$) also show 
the flipping with a very little change in the value of 
single-particle energy $\Delta\epsilon$.
In the analysis of neutron and proton single particle energy levels, 
we find that the trend of proton and neutron orbits are opposite 
to each other.
This nature gives rise to effect of change in neutron and 
proton radius in opposite trend. 

It is worthy to mention that the radius of $^{40}$Ca is slightly 
more than that of $^{48}$Ca (i.e. $r_{c}$=3.4776 of $^{40}$Ca 
and $r_{c}$=3.4771 of $^{48}$Ca~\cite{angeli2013,haddad2012}), 
and this is difficult to explain by most of the nuclear models.
We expact that similar anomaly may be occured in hyper-calcium 
($^{40}_\Lambda$Ca and $^{48}_\Lambda$Ca) also and can be solved by the 
additional $\delta-$meson degree of freedom to the model.
This mechanism can be used to solve the well known radius anomaly 
of $^{40}$Ca and $^{48}$Ca.

The neutron and lambda mean field potential 
for considered hypernuclei are plotted in Fig~\ref{fig7}.
The lambda central potential depth is found to be $V_\Lambda\sim$ 
32.87, 30.41 and 31.95 MeV for $^{48}_\Lambda$Ca, 
$^{90}_\Lambda$Zr and $^{208}_\Lambda$Pb, respectively.
It is to be noticed that the amount of lambda potential 
is 38$-$40\% of nucleon potential.
There are many of the calculations~\cite{keil2000,motoba1988,
davis1962} in prediction of lambda potential depth and our 
results are consistent with these predictions.
It is shown in Fig.~\ref{fig7} that both the potentials 
have similar shape but different depth.
It is also found that the lambda potential is completely 
unaffected with the strength of $\delta-$meson coupling.

\subsection{Spin-orbit splitting}
The spin-orbit interaction plays a crucial role in order
to investigate the structural properties of normal
as well as hypernuclei developed by the
exchange of scalar and vector mesons~\cite{koepf1991,brockmann1977}.
It is well known that the spin-orbit force in hypernuclei is weaker than
normal nuclear system~\cite{keil2000,brockmann1977,ajimura2001}.
Here, we study the spin-orbit potential for nucleon ($V_{so}^N$)
and hyperon ($V_{so}^\Lambda$) in hypernuclei and also analyze the effect 
of $g_\delta$ on spin-orbit interaction.
The spin-orbit potentials are displayed in Fig.~\ref{fig11} for 
considered hypernuclei.
To see the effect of $g_\delta$, we make a plot with $g_\delta$=0.0 
and for $g_\delta$=8.0, which is the largest allowed 
strength of delta-meson coupling. 
Figure~\ref{fig11} reveals that the spin-orbit potential for 
hyperons is weaker than their normal counter parts and these 
results are consistent with existing predictions
~\cite{keil2000,brockmann1977,ajimura2001}. 
It is clearly seen from the Fig.~\ref{fig11} that the delta-meson coupling 
does not have any valuable impact on spin-orbit interaction.
Actually, no change in spin-orbit potential is observed 
for $^{48}_\Lambda$Ca and $^{90}_\Lambda$Zr hypernuclei.
Rather than this, the spin-orbit potentials 
in $^{208}_\Lambda$Pb hypernucleus is affected by a very little amount.
This trend reflects that the measurable effect of $g_\delta$ on 
spin-orbit interaction can be observed 
from a system with large isospin asymmetry for example, 
heavy or superheavy nuclei and hypernuclei.


\section{SUMMARY AND CONCLUSIONS}
In summary, we study the contribution as well as importance 
of $\delta-$meson coupling in non-linear RMF model for hypernuclei.
The lambda potential depth is found to be consistent with 
other predictions~\cite{keil2000,motoba1988,davis1962}.
The calculated $B_\Lambda$ for considered nuclei are quite 
agreeable with the experimental data. 
In the present calculation, we have included it to reveal 
the effects of $g_\delta$ coupling strength on hypernuclei 
which are found to be significant.
It is clear to say that $g_\delta$ affects every piece of 
physical observables of hypernuclei, like binding energy, 
radii, single particle energy and spin-orbit splitting 
for nuclear system with N$\neq$Z, but the magnitude of 
affects is less comparable to normal nuclei.
Contrary to this, the proton and charge radii are affected 
relatively more than normal nuclear case.
A very small reduction in lambda radius is also observed with 
increasing strength of $\delta-$meson coupling.
However, the lambda potential is completely unaffected 
by $\delta-$meson coupling strength due to zero isospin 
nature of $\Lambda$ particle.  
The variation of spin-orbit interaction is discussed in  
respect of $\delta$-meson coupling.
This coupling does not have any significant impact on spin-orbit 
potential for considered hypernuclei but reflects that its 
impact would be measurable for a system with large isospin asymmetry.
It is clearly seen that the contribution of $\delta-$meson 
is more effective with the magnitude of asymmetry of the system.
From the given results, it is concluded that $\delta-$meson 
has indispensable contribution not only in asymmetric nuclei 
but also for hypernuclei.

The $\delta-$meson coupling may prove to be a significant degree of 
freedom for resolving the charge radius anomaly which is 
appeared in $^{40}$Ca and $^{48}$Ca and also if happened 
in corresponding hypernuclei.
The production of $^{48}_\Lambda$Ca hypernucleus is possible in 
future due to advanced experimental facilities across the world.

\section*{ACKNOWLEDGMENTS}
One of the author (MI) would like to acknowledge the hospitality 
provided by Institute of Physics, Bhubaneswar during the work.


\end{document}